# Reflection and Transmission Coefficients of the symmetric barrier-type shifted Deng-Fan potential

O J Oluwadare[+1], K J Oyewumi[+2], K-E Thylwe[+3] and S Hassanabadi[+4]

[+1]Department of Physics, Federal University Oye-Ekiti, P. M. B. 373, Oye-Ekiti, Ekiti State, Nigeria
[+2]Theoretical Physics Section, Department of Physics, University of Ilorin, P. M. B. 1515, Ilorin, Nigeria
[+3]KTH-Mechanics, Royal Institute of Technology, S-100 44 Stockholm, Sweden
[+4]Physics Department, Semnan University, Semnan, Iran

*Abstract: By applying continuity and boundary conditions, the reflection and transmission coefficients of one-dimensional time-independent Schrödinger equation with a symmetric* barrier-type *shifted Deng-Fan potential are obtained and discussed. The numerical and graphical results are very accurate and consistent with the quantum transmission-reflection principle.*

## 1. Introduction

The scattering phenomena, such as scattering phase shifts, reflection and transmission coefficients are used to explain the behaviour of waves incident on a barrier [1-2]. The transmission coefficient represents the probability flux of the transmitted wave relative to that of the incident wave. It is often used to describe the probability of a particle tunneling through a barrier. This is more feasible in the case where electrons tunnel between two conducting materials. The shifted Deng-Fan potential is a suitable model for the theoretical predictions because it has been used extensively in both relativistic and non-relativistic quantum mechanics. The original Deng-Fan molecular potential has the correct asymptotic behaviour at the origin [3-9], and so, it has been used to predict the interactions between atoms and molecules and vibrational spectrum [3-9].

Just of recent, a modified form of Deng-Fan potential called the shifted Deng-Fan potential was suggested by Hamzavi et al. (2013), it was reported that the shifted Deng-Fan potential and the Morse potential are very close to each other for large values of $x$ in the domain where $x \approx x_e$ and $x > x_e$ but different at $x \approx 0$ [10]. Yahya et al. (2013) obtained the solutions of the Dirac equation with the shifted Deng-Fan potential including Yukawa-like tensor interaction [11].

Oyewumi et al. (2014) studied the thermodynamic properties of the Schrödinger equation with the shifted Deng-Fan potential [12]. Roy (2014) considered the ro-vibrational studies of diatomic molecules in a shifted Deng-Fan oscillator potential by applying the pseudo-spectral method which encouraged Mustafa (2015) to investigate a new deformed Schioberg-type potential, and obtained ro-vibrational energies for some diatomic molecule [13]. He showed how this potential is related to the Deng-Fan potential [14]. Thus, the study of reflection and transmission coefficients of the symmetric barrier-type shifted Deng-Fan potential model is very necessary.

Very recently, attention has been directed to the studies on barrier penetration problems. Villalba & Greiner (2003) investigated the transmission resonances and supercritical states by solving the two-component Dirac equation for the Cusp potential [15]. Thylwe (2005) studied the non-relativistic barrier transmission problem treated by the amplitude-phase method and expressed in terms of an invariant of the Ermakov-Lewis type [16]. He presented the analysis of the transmission and reflection of a 'quantal' by a single-hump potential barrier and concluded that these coefficients are functions of Ermakov-Lewis type invariants. Arda et al. (2011) studied the scattering and bound state solutions of asymmetric Hulthén potential and presented the reflection and transmission coefficients of the one-dimensional Schrödinger equation [17].

In view of a spatially one-dimensional asymmetric Hulthén potential, Hassanabadi et al. [18] looked into the two-body scattering in (1+1) dimension by a semi-relativistic formalism and a piecewise Hulthén interaction potential. The reflection and transmission coefficients were calculated and discussed.

Consequently, Thylwe et al. (2016) investigated semi-relativistic reflection and transmission coefficients for two spinless particles separated by exponential- and rectangular-shaped potential barriers. They obtained the reflection and transmission coefficients by the non-relativistic (Schrödinger-) approximation and also presented the approximate quantum mechanical two-body analyses of spinless particles, scattered by an inter-particle barrier in 1+1 dimensions [19].

The present study focuses on the reflection and transmission coefficients of non-relativistic particles incident on the one dimensional piecewise shifted Deng-Fan type potential barrier. The exact results for this potential are reliable and consistent within the non-relativistic Schrödinger equation.

The present work is organized as follows. In Section 2, we present the basic one-dimensional time-independent Schrodinger equation for a particle with mass $m$ approaching a potential $V(x)$. In Section 3, we obtain the reflection and transmission coefficients of Schrödinger wave equation with a symmetric 'barrier-type' shifted Deng-Fan potential. Conclusions are given in Section 4.

## 2. The Basic Equation

The one-dimensional time-independent Schrodinger equation for a particle with mass $m$ approaching a potential $V(x)$ is given by

$$\psi''(x) + 2m[E - V(x)]\psi(x) = 0. \tag{1}$$

The one dimensional symmetric 'barrier-type' shifted Deng-Fan potential studied is given [10-14] by

$$V(x) = V_0 b \left[ \frac{b}{(e^{a|x|}-q)^2} - \frac{2}{(e^{a|x|}-q)} \right], \quad b = e^{ax_e} - q, \tag{2}$$

where $x_e$ is the equilibrium distance, $V_0$ is dissociation energy, $a$ is the inverse potential range and the positive parameter $q < 1$ hold at the origin.

## 3. The Reflection and Transmission Coefficients

By considering the symmetric shifted Deng-Fan potential for the region $x < 0$, Eq. (1) yields

$$\psi_L''(x) + 2m \left\{ E - V_0 b \left[ \frac{b}{(e^{-ax}-q)^2} - \frac{2}{(e^{-ax}-q)} \right] \right\} \psi_L(x) = 0. \tag{3}$$

As $x \to -\infty$ in the region, with the aid of a new variable $y_L = qe^{ax}$ in Eq. (3), we obtain the following linear differential equation:

$$y_L(1-y_L)\psi_L''(y_L) + (1-y_L)\psi_L'(y_L) + \frac{1}{y_L(1-y_L)}[\chi_1 y_L^2 + \chi_2 y_L + \chi_3]\psi_L(y_L) = 0, \tag{4}$$

where

$$-\chi_1 = -\frac{2mE}{a^2} + \frac{2mV_0 b^2}{a^2 q^2} + \frac{4mV_0 b}{a^2 q}; \quad \chi_2 = \frac{4mV_0 b}{a^2 q} - \frac{4mE}{a^2}; \quad \chi_3 = \frac{2mE}{a^2}. \tag{5}$$

To transform Eq. (4) into equation of hypergeometric type, we employed a trial wave function of the type:

$$\psi_L(y_L) = y_L^{\sigma_L}(1-y_L)^{\tau_L} f(y_L), \tag{6}$$

and with the aid of this in Eq. (4), we obtain the equation of hypergeometric type as:

$$f''(y_L) + \frac{[1+2\sigma_L - (2\sigma_L + 2\tau_L + 1)y_L]}{y_L(1-y_L)} f'(y_L) + \frac{[\sigma_L^2 + \tau_L^2 + 2\sigma_L \tau_L + \chi_1]}{y_L(1-y_L)} f(y_L) = 0. \tag{7}$$

As $x \to -\infty$ in the region $(x < 0)$ the wave function is the sum of the incident wave and the reflected wave and is of the type [17, 18, 20, 21]

$$\psi_L(y_L) = A_1 y_L^{\sigma_L}(1-y_L)^{\tau_L} {}_2F_1(\alpha, \beta; \gamma; y_L)$$

$$+A_2 y_L^{-\sigma_L}(1-y_L)^{\tau_L}{}_2F_1(\alpha+1-\gamma,\beta+1-\gamma;2-\gamma;y_L), \quad (8)$$

where we have used the following parameters for simplicity:

$$\alpha = \sigma_L + \tau_L - \sqrt{-\chi_1}, \tag{9a}$$

$$\beta = \sigma_L + \tau_L + \sqrt{-\chi_1}, \tag{9b}$$

$$\gamma = 1 + 2\sigma_L, \tag{9c}$$

$$\epsilon_L = \chi_1 + \chi_2 + \chi_3, \tag{9d}$$

$$\tau_L = \frac{1}{2} \pm \frac{1}{2}\sqrt{1-4\epsilon_L}. \tag{9e}$$

It is worth nothing that $\sigma_L = \frac{ik}{a}$, and that $k = \sqrt{2mE}$ is the asymptotic wave number.

By considering the symmetric potential for the region $x > 0$, for which the positive parameter $\tilde{q} < 1$ holds, Eq. (1) yields

$$\psi_R''(x) + 2m\left\{E - V_0 b\left[\frac{b}{(e^{ax}-\tilde{q})^2} - \frac{2}{(e^{ax}-\tilde{q})}\right]\right\}\psi_R(x) = 0. \tag{10}$$

To obtain a solution as $x \to \infty$ in the region ($x > 0$), we apply a new variable $y_R = \tilde{q}e^{-ax}$ in Eq. (10), yielding

$$y_R(1-y_R)\psi_R''(y_R) + (1-y_R)\psi_R'(y_L) + \frac{1}{y_R(1-y_R)}[\chi_4 y_R^2 + \chi_5 y_R + \chi_6]\psi_R(y_R) = 0, \tag{11}$$

where

$$-\chi_4 = -\frac{2mE}{a^2} + \frac{2mV_0 b^2}{a^2 \tilde{q}^2} + \frac{4mV_0 b}{a^2 \tilde{q}}; \quad \chi_5 = \frac{4mV_0 b}{a^2 \tilde{q}} - \frac{4mE}{a^2}; \quad \chi_6 = \frac{2mE}{a^2}. \tag{12}$$

Transforming Eq. (10) into equation of hypergeometric type is done by employing a similar trial wave function of the type:

$$\psi_R(y_R) = y_R^{\sigma_R}(1-y_R)^{\tau_R} f(y_R), \tag{13}$$

which yields another equation of hypergeometric type of the form:

$$f''(y_R) + \frac{[1+2\sigma_R - (2\sigma_R + 2\tau_R + 1)y_R]}{y_R(1-y_R)} f'(y_R) + \frac{[\sigma_R^2 + \tau_R^2 + 2\sigma_R \tau_R + \chi_4]}{y_R(1-y_R)} f(y_R) = 0. \tag{14}$$

As $x \to \infty$ in the region ($x > 0$), the wave function is the transmitted wave only and is of the type [17, 18, 20, 21]

$$\psi_R(y_R) = A_3 y_R^{\sigma_R}(1-y_R)^{\tau_R}{}_2F_1(\tilde{\alpha},\tilde{\beta};\tilde{\gamma};y_R)$$

$$+A_4 y_R^{-\sigma_R}(1-y_R)^{\tau_R}{}_2F_1(\tilde{\alpha}+1-\tilde{\gamma},\tilde{\beta}+1-\tilde{\gamma};2-\tilde{\gamma};y_R), \tag{15}$$

where we have used the following parameters for simplicity:

$$\tilde{\alpha} = \sigma_R + \tau_R - \sqrt{-\chi_4}, \tag{16a}$$

$$\tilde{\beta} = \sigma_R + \tau_R + \sqrt{-\chi_4}, \tag{16b}$$

$$\tilde{\gamma} = 1 + 2\sigma_R, \tag{16c}$$

$$\epsilon_R = \chi_4 + \chi_5 + \chi_6, \tag{16d}$$

$$\tau_R = \frac{1}{2} \pm \frac{1}{2}\sqrt{1 - 4\epsilon_R}. \tag{16e}$$

Again $\sigma_R = \frac{ik}{a}$, and $k = \sqrt{2mE}$ is the asymptotic wave number. Since no wave is reflected from the region $x > 0$, we thus set the constant $A_3$ to zero in Eq. (15) and have [17, 18, 20, 21]

$$\psi_R(y_R) = A_4 y_R^{-\sigma_R}(1 - y_R)^{\tau_R} {}_2F_1(\tilde{\alpha} + 1 - \tilde{\gamma}, \tilde{\beta} + 1 - \tilde{\gamma}; 2 - \tilde{\gamma}; y_R). \tag{17}$$

Now, we consider the asymptotic behaviour of the wave functions on the both sides, i.e.

$$x \to -\infty, y_L \to 0, (1 - y_L)^{\tau_L} \to 1, y_L^{\pm\sigma_L} \to q^{\pm\sigma_L} e^{\pm ax\sigma_L}, \tag{18a}$$

$$x \to \infty, y_R \to 0, (1 - y_R)^{\tau_R} \to 1, y_R^{\mp\sigma_R} \to \tilde{q}^{\mp\sigma_R} e^{\pm ax\sigma_R}. \tag{18b}$$

Consequently, we summarize the total wave function for the limit $x \to \pm\infty$ as:

$$\psi(x) = \begin{cases} A_1 q^{\sigma_L} e^{ax\sigma_L} + A_2 q^{-\sigma_L} e^{-ax\sigma_L}, & x \to -\infty, \\ A_4 \tilde{q}^{-\sigma_R} e^{ax\sigma_R}, & x \to -\infty. \end{cases} \tag{19}$$

In order to evaluate the explicit expressions for the coefficients $A_1$, $A_2$, and $A_4$ employed in Eq. (8) and Eq. (17), we need to apply the continuity or boundary conditions on the wave function and their derivatives, i.e. $\psi_L(x = 0) = \psi_R(x = 0)$ and $\psi_L'(x = 0) = \psi_R'(x = 0)$ which respectively, give the following equations:

$$A_1 \rho_1^{\sigma_L} \rho_2^{\tau_L} \zeta_1 + A_2 \rho_1^{-\sigma_L} \rho_2^{\tau_L} \zeta_2 = A_4 \rho_3^{-\sigma_R} \rho_4^{\tau_R} \zeta_3 \tag{20}$$

and

$$A_1 \left[\sigma_L \rho_1^{\sigma_L-1} \rho_2^{\tau_L} \zeta_1 - \tau_L \rho_1^{\sigma_L} \rho_2^{\tau_L-1} \zeta_1 + \rho_1^{\sigma_L} \rho_2^{\tau_L} \Lambda_1 \zeta_4\right] + A_2 \left[-\sigma_L \rho_1^{-\sigma_L-1} \rho_2^{\tau_L} \zeta_2 - \tau_L \rho_1^{-\sigma_L} \rho_2^{\tau_L-1} \zeta_2 + \rho_1^{-\sigma_L} \rho_2^{\tau_L} \Lambda_2 \zeta_5\right] = A_4 \left[-\sigma_R \rho_3^{-\sigma_R-1} \rho_4^{\tau_R} \zeta_3 - \tau_R \rho_3^{-\sigma_R} \rho_4^{\tau_R-1} \zeta_3 + \rho_3^{-\sigma_R} \rho_4^{\tau_R} \Lambda_3 \zeta_6\right], \tag{21}$$

where we have employed useful parameters in the process of derivation.

$$\rho_1 = q, \quad \rho_2 = 1 - q, \quad \rho_3 = \tilde{q}, \quad \rho_4 = 1 - \tilde{q} \tag{22}$$

$$\zeta_1 = {}_2F_1(\alpha, \beta; \gamma; \rho_1), \tag{23a}$$

$$\zeta_2 = {}_2F_1(\alpha + 1 - \gamma, \beta + 1 - \gamma; 2 - \gamma; \rho_1), \tag{23b}$$

$$\zeta_3 = {}_2F_1(\tilde{\alpha} + 1 - \tilde{\gamma}, \tilde{\beta} + 1 - \tilde{\gamma}; 2 - \tilde{\gamma}; \rho_3), \tag{23c}$$

$$\zeta_4 = {}_2F_1(\alpha + 1, \beta + 1; \gamma + 1; \rho_1), \tag{23d}$$

$$\zeta_5 = {}_2F_1(\alpha + 2 - \gamma, \beta + 2 - \gamma; 3 - \gamma; \rho_1), \tag{23e}$$

$$\zeta_6 = {}_2F_1(\tilde{\alpha} + 1 - \tilde{\gamma}, \tilde{\beta} + 1 - \tilde{\gamma}; 2 - \tilde{\gamma}; \rho_3), \tag{23f}$$

$$\Lambda_1 = \frac{(\alpha)(\beta)}{\gamma}, \Lambda_2 = \frac{(\alpha+1-\gamma)(\beta+1-\gamma)}{2-\gamma}, \Lambda_3 = \frac{(\tilde{\alpha}+1-\tilde{\gamma})(\tilde{\beta}+1-\tilde{\gamma})}{2-\tilde{\gamma}}. \tag{23g}$$

The Reflection coefficient $R$ and the transmission coefficient $T$ are defined [17, 18], respectively as

$$R = \left|\frac{A_2}{A_1}\right|^2 = \left|\frac{c_3 c_4 - c_1 c_6}{c_2 c_6 - c_3 c_5}\right|^2, \tag{24}$$

$$T = \left|\frac{A_4}{A_1}\right|^2 = \left|\frac{c_2 c_4 - c_1 c_5}{c_2 c_6 - c_3 c_5}\right|^2, \tag{25}$$

where

$$c_1 = \rho_1^{\sigma_L} \rho_2^{\tau_L} \zeta_1, \; c_2 = \rho_1^{-\sigma_L} \rho_2^{\tau_L} \zeta_2, \; c_3 = \rho_3^{-\sigma_R} \rho_4^{\tau_R} \zeta_3, \tag{26a}$$

$$c_4 = \sigma_L \rho_1^{\sigma_L-1} \rho_2^{\tau_L} \zeta_1 - \tau_L \rho_1^{\sigma_L} \rho_2^{\tau_L-1} \zeta_1 + \rho_1^{\sigma_L} \rho_2^{\tau_L} \Lambda_1 \zeta_4, \tag{26b}$$

$$c_5 = -\sigma_L \rho_1^{-\sigma_L-1} \rho_2^{\tau_L} \zeta_2 - \tau_L \rho_1^{-\sigma_L} \rho_2^{\tau_L-1} \zeta_2 + \rho_1^{-\sigma_L} \rho_2^{\tau_L} \Lambda_2 \zeta_5, \tag{26c}$$

$$c_6 = -\sigma_R \rho_3^{-\sigma_R-1} \rho_4^{\tau_R} \zeta_3 - \tau_R \rho_3^{-\sigma_R} \rho_4^{\tau_R-1} \zeta_3 + \rho_3^{-\sigma_R} \rho_4^{\tau_R} \Lambda_3 \zeta_6. \tag{26d}$$

## 3.1 Numerical Results

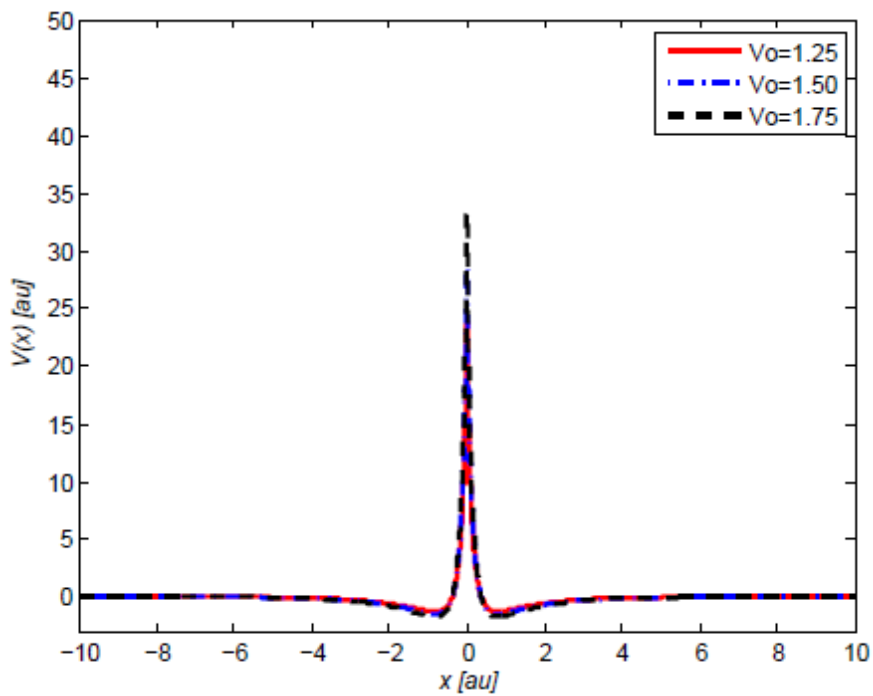

Figure 1: The plot of shifted Deng-Fan potential $V(x)$ versus $x$ for the different values of $V_0$ and the parameters $a = 0.8$, $x_e = 0.8$ and $q = \tilde{q} = 0.8$.

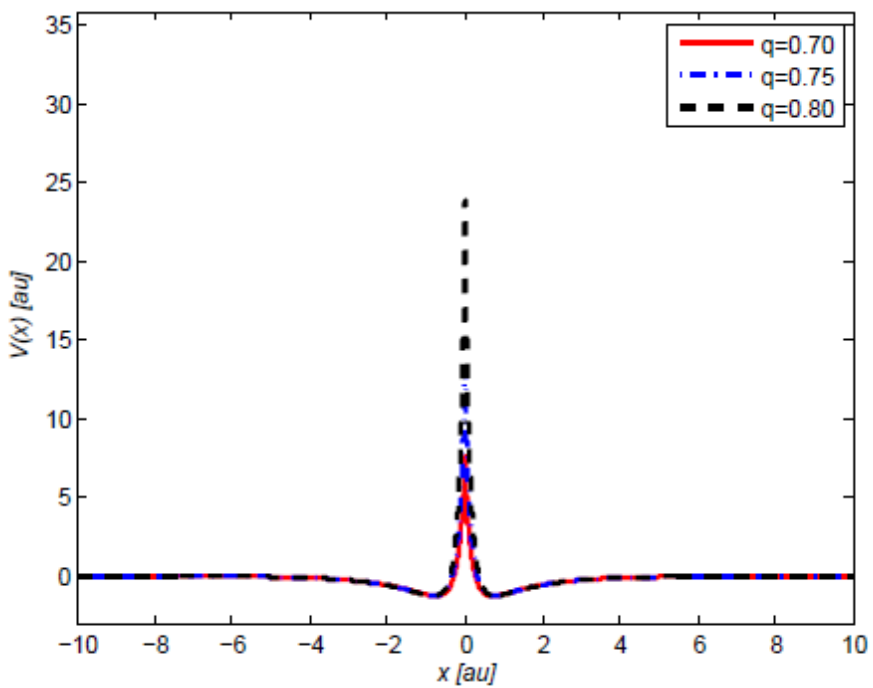

Figure 2: The plot of shifted Deng-Fan potential $V(x)$ versus $x$ for the different values of $q = \tilde{q}$ and the parameters $a = 0.8$, $x_e = 0.8$ and $V_0 = 1.25$

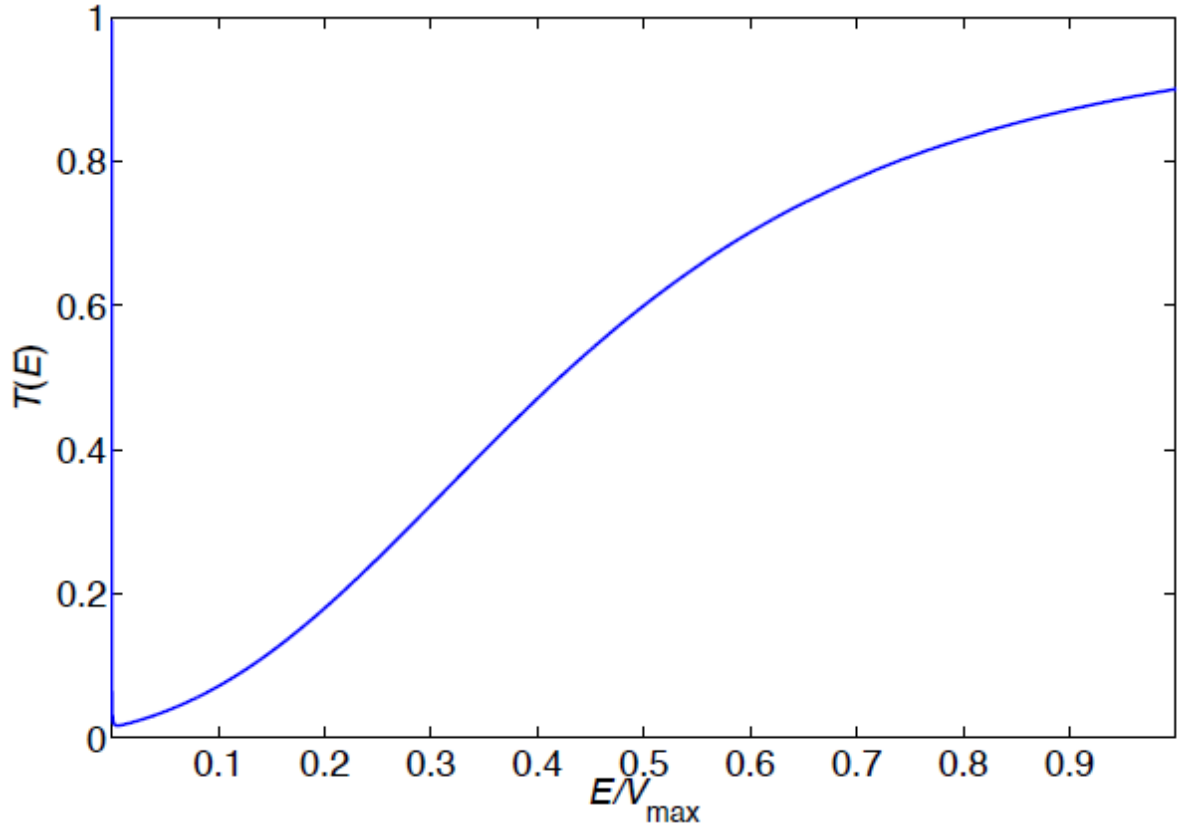

Figure 3: The plot of transmission coefficient $T(E)$ versus $E/V_{max}$, where $V_{max} = V(0)$, for different values of and the parameters $a = x_e = q = \tilde{q} = 0.8$, $m = 1$ and $V_0 = 1.25$.

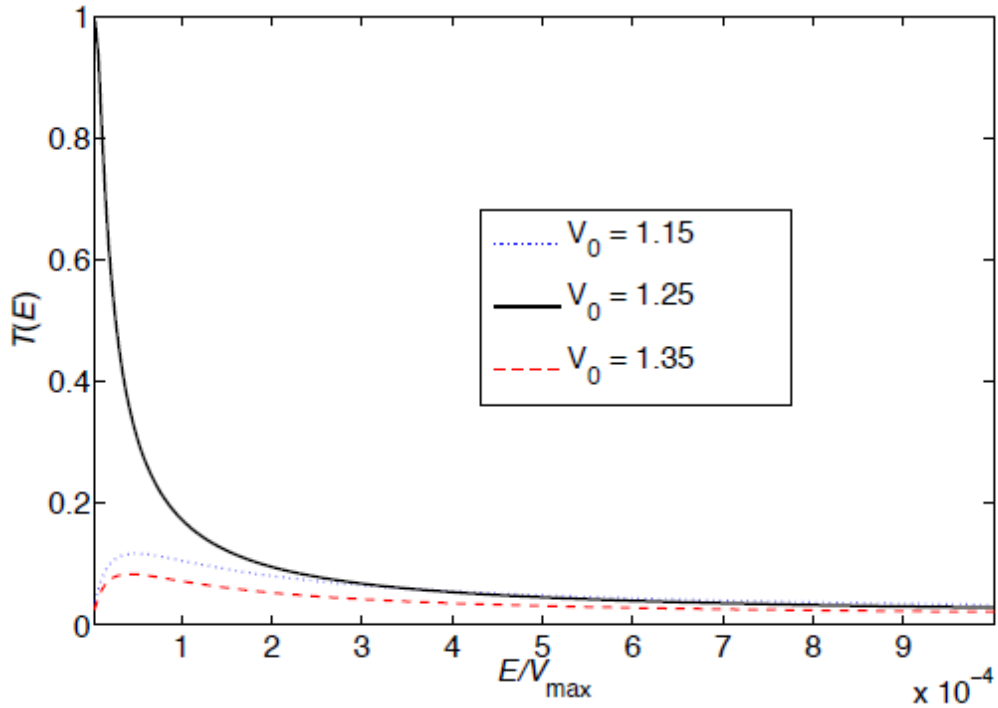

Figure 4: The plot of transmission coefficient $T(E)$ versus $E/V_{max}$, where $V_{max} = V(0)$, for different values of $V_0$ and the parameters $a = x_e = q = \tilde{q} = 0.8$ and $m = 1$.

**Table 1**: The Reflection and Transmission coefficients as a function of selected values of Energies $E$ for the parameters $V_0 = 1.25, a = x_e = q = \tilde{q} = 0.8$ and $m = 1$.

| $E$ | $T(E)$ | $R(E)$ |
|---|---|---|
| 0.005 | 0.0992153 | 0.900785 |
| 0.010 | 0.0559170 | 0.944083 |
| 0.015 | 0.0411413 | 0.958859 |
| 0.020 | 0.0337481 | 0.966252 |
| 0.025 | 0.0293473 | 0.970653 |
| 0.030 | 0.0264526 | 0.973547 |
| 0.035 | 0.0244214 | 0.975579 |
| 0.040 | 0.0229305 | 0.977069 |
| 0.045 | 0.0217998 | 0.978200 |
| 0.050 | 0.0209209 | 0.979079 |
| 0.055 | 0.0202247 | 0.979775 |
| 0.060 | 0.0196651 | 0.980335 |
| 0.065 | 0.0192101 | 0.980790 |
| 0.070 | 0.0188371 | 0.981163 |
| 0.075 | 0.0185293 | 0.981471 |
| 0.080 | 0.0182742 | 0.981726 |
| 0.085 | 0.0180621 | 0.981938 |
| 0.090 | 0.0178858 | 0.982114 |
| 0.095 | 0.0177393 | 0.982261 |
| 0.100 | 0.0176180 | 0.982382 |

Figures 1 and 2 illustrate the potential barrier for selected sets of the potential parameters $V_0$ and $q = \tilde{q}$. $V_0$ corresponds to the depth of the well and large values make the well deeper and barrier maximum larger. A decrease of the parameter $q = \tilde{q}$ ($0 \leq q = \tilde{q} \leq 1$) causes a shift of the right-hand potential shape (particularly the well) to the left so that the barrier maximum at the origin decreases. A symmetric behaviour results for the left-hand potential shape. The barrier in this study is relatively high ($V(0) = V_{max}$) in comparison with its width.

Figure 3 shows a typical large-scale behaviour of the transmission coefficient as function of the energy relative to the barrier top energy $V_{max}$. It tends to unity far beyond the energy of the barrier top. We observe a tiny detail that the transmission coefficient does not tend to zero as the scattering energy tends to zero.

The low-energy behaviour is studied in more detail in Figure 4. Figure 4 shows low-energy behaviour of the transmission coefficient T for $V_0$ =1.15, 1.25 and 1.35, with $a = x_e = q = \tilde{q} = 0.8$ and m=1 in atomic units. The reflection coefficient R is obtained from the exact relation T+R=1. While for $V_0$ =1.25 there is almost total transmission at zero energy and there are local low-energy peaks of transmission for $V_0$ =1.15 and 1.35.

Table 1 presents accurate values of the transmission coefficient T for selected values of the energy. The results have been recalculated and confirmed with the aid of the amplitude-phase method [15].

## 4    Conclusion

We have investigated the one-dimensional time-independent Schrödinger equation with a symmetric barrier surrounded by shallow wells on both sides of the barrier. The shape on either side of the barrier top is that of a shifted Deng-Fan potential which has been cut at the origin. Continuity and scattering boundary conditions are applied in a standard way. After transforming the Schrödinger equation to a suitable form having hypergeometric solutions, we derive the expressions for the reflection (R) and transmission (T) coefficients satisfying T+R=1.

Numerical results are exact, except for truncations, and have been verified by an independent numerical amplitude-phase method [15]. The presence of a resonance phenomenon is found for a particular set of barrier parameters, which allows total transmission at zero scattering energy and transmission peaks at energies close to zero.

We conclude that the shallow well is responsible for the resonance phenomenon and that the transformation method produce exact results for the particular potential barrier used in the present work.